\documentclass[prl,showpacs,amsfonts,amsmath,twocolumn,superscriptaddress]{revtex4-1}
\pdfoutput=1
\usepackage{array,amsmath,amsfonts,amssymb,dsfont}
\usepackage[T1]{fontenc}
\usepackage{ae}
\usepackage{bbold}
\usepackage{graphicx}
\usepackage{times}
\usepackage{natbib}
\newcommand{\ket}[1]{|#1\rangle}

\begin{document}

\title{Photon-mediated interactions between distant artificial atoms}

\author{A.~F.~van~Loo}
 \email{arjan@phys.ethz.ch}
 \affiliation{Department of Physics, ETH Zurich, CH-8093 Zurich, Switzerland}
\author{A.~Fedorov}
\altaffiliation[Present address: ]{ARC Centre for Engineered Quantum Systems, University of Queensland, Brisbane 4072, Australia}
\affiliation{Department of Physics, ETH Zurich, CH-8093 Zurich, Switzerland}
\author{K.~Lalumi\`ere}
\affiliation{D\'{e}partement de Physique, Universit\'e de Sherbrooke, Sherbrooke, Qu\'{e}bec J1K 2R1, Canada}
\author{B.~C.~Sanders}
 \affiliation{Institute for Quantum Science and Technology, University of Calgary, Alberta T2N 1N4, Canada}
\author{A.~Blais}
\affiliation{D\'{e}partement de Physique, Universit\'e de Sherbrooke, Sherbrooke, Qu\'{e}bec J1K 2R1, Canada}
\author{A.~Wallraff}
 \affiliation{Department of Physics, ETH Zurich, CH-8093 Zurich, Switzerland}

%%%%%%%%%%%%%%%%% END OF PREAMBLE %%%%%%%%%%%%%%%%

\begin{abstract}
Photon-mediated interactions between atoms are of fundamental importance in quantum optics, quantum simulations and quantum information processing. The exchange of real and virtual photons between atoms gives rise to non-trivial interactions the strength of which decreases rapidly with distance in three dimensions. Here we study much stronger photon mediated interactions using two superconducting qubits in an open one-dimensional transmission line. Making use of the unique possibility to tune these qubits by more than a quarter of their transition frequency we observe both coherent exchange interactions at an effective separation of $3\lambda/4$ and the creation of super- and sub-radiant states at a separation of one photon wavelength $\lambda$. This system is highly suitable for exploring collective atom/photon interactions and applications in quantum communication technology.
\end{abstract}

\maketitle

In free space the interaction of individual atoms with vacuum fluctuations of the electromagnetic field leads to both the relaxation of atomic excited states and the renormalization of atomic energy levels. On one hand, the emission of real photons into a single mode of the electromagnetic continuum at the atomic transition frequency results in spontaneous emission. On the other hand the emission and absorption of virtual photons from all modes of the continuum gives rise to a Lamb shift (a shift in the energy level separation). In the presence of additional atoms, both real and virtual photons emitted by one atom can be absorbed by another one, giving rise to non-trivial atom-atom interactions. Such interactions are challenging to observe in three dimensions due to the weak electromagnetic fields generated by individual photons and the poor spatial mode matching between the modes of the emitting and absorbing atoms. Nevertheless, signatures of these interactions the form of super- and sub-radiant states, which depended on the separation of two trapped ions, were observed~\cite{DeVoe1996,Eschner2001a}. Due to the inverse scaling of the interaction strength with inter-atomic separation the super- and sub-radiant lifetimes were observed to differ by only a few percent in these experiments.

Confining both the radiation field and the two-level systems to one dimension overcomes the aforementioned challenges and allows for observing photon-mediated interactions between atoms. This nascent field of physics known as waveguide quantum electrodynamics (QED) is expected to contribute to the development of quantum networks~\cite{Kimble2008}, circuits operating on the level of single photons~\cite{Chang2007,Hwang2009}, implementations of quantum memories using EIT~\cite{Leung2012a} and the generation of photon-photon interactions~\cite{Zheng2012h}.

At optical frequencies, realizations of waveguide QED systems have been proposed for quantum dots interacting with surface plasmons~\cite{Chang2006,Chang2007a,Akimov2007,Martin-Cano2010,Gonzalez-Tudela2011} and for atoms trapped in the near-field of a nanofiber~\cite{Nayak2007,Vetsch2010,Chang2012a,Goban2012}. Superconducting circuits are also a natural choice to investigate the strong interaction of quantum two-level systems with microwave photons in open one-dimensional (1D) transmission lines~\cite{Shen2005}. Near-unit reflectance of weak coherent fields by a single qubit was first observed by Astafiev \textit{et. al.}~\cite{Astafiev2010}. Subsequently, phenomena such as resonance fluorescence, Autler-Townes splitting, electromagnetically induced transparency~\cite{Abdumalikov2010}, time-resolved emission dynamics~\cite{Abdumalikov2011}, and the cross-Kerr effect~\cite{Hoi2013b} have been explored, and single-photon routers~\cite{Hoi2011} have been realized.

Here we demonstrate the coupling between two superconducting qubits mediated by microwave photons in a 1D transmission line. In contrast to the 3D case in which the interaction decreases rapidly with increasing separation between the qubits, in 1D it shows an oscillating behavior at an approximately constant amplitude (see Ref.~\cite{Lalumiere2013} and references therein). Indeed, the photon-mediated interaction leads to correlated decay of a pair of qubits at a rate $\gamma \propto \cos (2\pi d/\lambda)$ and coherent exchange-type interactions at a frequency $J \propto \sin (2\pi d/\lambda)$. Whereas the inter-qubit separation $d$ is fixed for a given sample in our experiment, we vary the effective separation between the qubits in terms of their emission wavelength $\lambda$. For this purpose we change the qubit transition frequencies by an appreciable fraction of their maximal values (Fig.~1A), an aspect which is challenging to achieve in most atomic systems.

\begin{figure*}[!t]
  \centering
    \includegraphics[width=12.0cm]{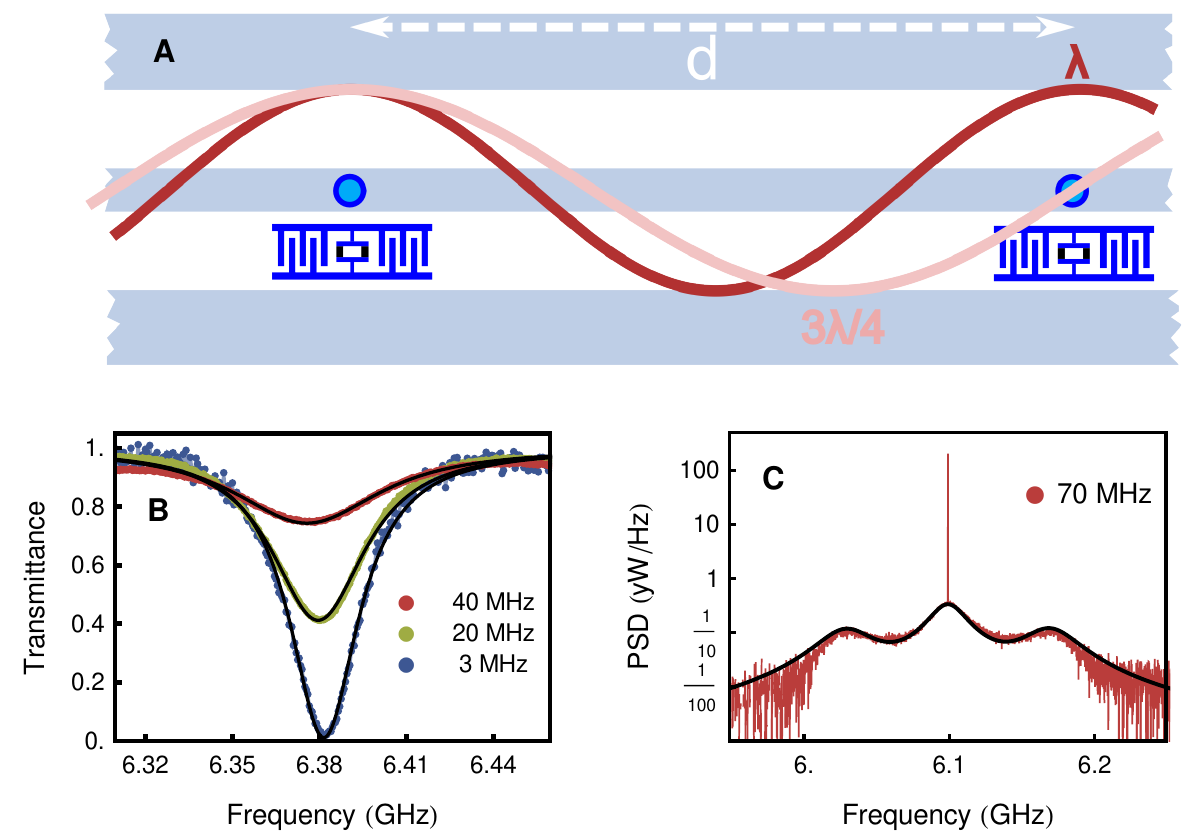}
    \caption{\textbf{The experimental system.} (A) Schematic of two transmon qubits coupled to an open transmission line at a fixed separation $d$ corresponding to an effective separation of $\lambda$ or $3\lambda/4$ which is tunable by adjusting the qubit transition frequency. (B) Transmittance spectrum $|t|^2$ of a single qubit measured at the indicated drive rates. (C) Power spectral density (PSD) of the radiation reflected by a single qubit. Red lines are data; black lines denote theory, see text for details.}
  \label{fig:setup}
\end{figure*}

We have fabricated a sample in which two superconducting transmon qubits~\cite{Koch2007} are coupled to a 1D coplanar waveguide transmission line at an inter-qubit separation of $d = 18.6 \, \rm{mm}$ (Fig.~1A). For both qubits we have determined the maximum frequency $6.89\,\rm{GHz}$ and $6.84\,\rm{GHz}$ of the ground $\ket{g}$ to first excited state $\ket{e}$ transitions and the identical anharmonicity $-298\,\rm{MHz}$ of the first to second excited state transition using spectroscopic techniques.

To further characterize the system we measured the amplitude and phase of a weak drive field coherently scattered from the sample in a detection bandwidth much less than  $1\,\rm{MHz}$~ (see the supplementary materials). We determined the transmittance $|t|^2$ and reflectance $|r|^2$ (normalized transmitted and reflected power) from the measured transmission and reflection coefficients $t$ and $r$ (see the supplementary materials). When either qubit is detuned by many linewidths from the other, we observed no difference in lineshape compared to a single qubit. At 6.4~GHz, the minimum transmittance is less than 0.025 for low drive powers (Fig.~1B), indicating strong coupling (see supplementary materials).

From the linewidth, the decay rates of both qubits are determined to be $\gamma_1/2\pi \approx 26\pm1\,\rm{MHz}$ at $6.4\,\rm{GHz}$, and $13\pm1\,\rm{MHz}$ at $4.8\,\rm{GHz}$, consistent with expectations for an ohmic environment~\cite{Astafiev2010}. When the power of the incident drive-field is increased, the qubit transition saturates \cite{Hoi2012b}, and transmittance increases to unity for large drive powers (Fig.~1B).

When the two qubits are tuned into resonance, such that a photon emitted by one of the qubits may be absorbed by the other, we expect correlated effects to become apparent. For both qubits tuned to either $6.4$ or $4.8\,\rm{GHz}$, corresponding to an effective qubit separation $d$ of $\lambda$ or $3\lambda/4$, respectively, we tune the qubits through resonance and plot $|t|^2$ and $|r|^2$ versus frequency (Fig.~2). When the qubits are detuned from each other by much more than the range displayed in Fig.~2 they each display a pronounced well-resolved maximum in $|r|^2$, similar to data obtained for a single qubit. When the two qubits are in resonance with each other for $d\sim\lambda$, only a single resonance is observed with $|r|^2$ reaching unity at mutual resonance and $|t|^2$ going to zero (Fig.~2B).

\begin{figure*}[!t]
  \begin{center}
    \includegraphics[width=12.0cm]{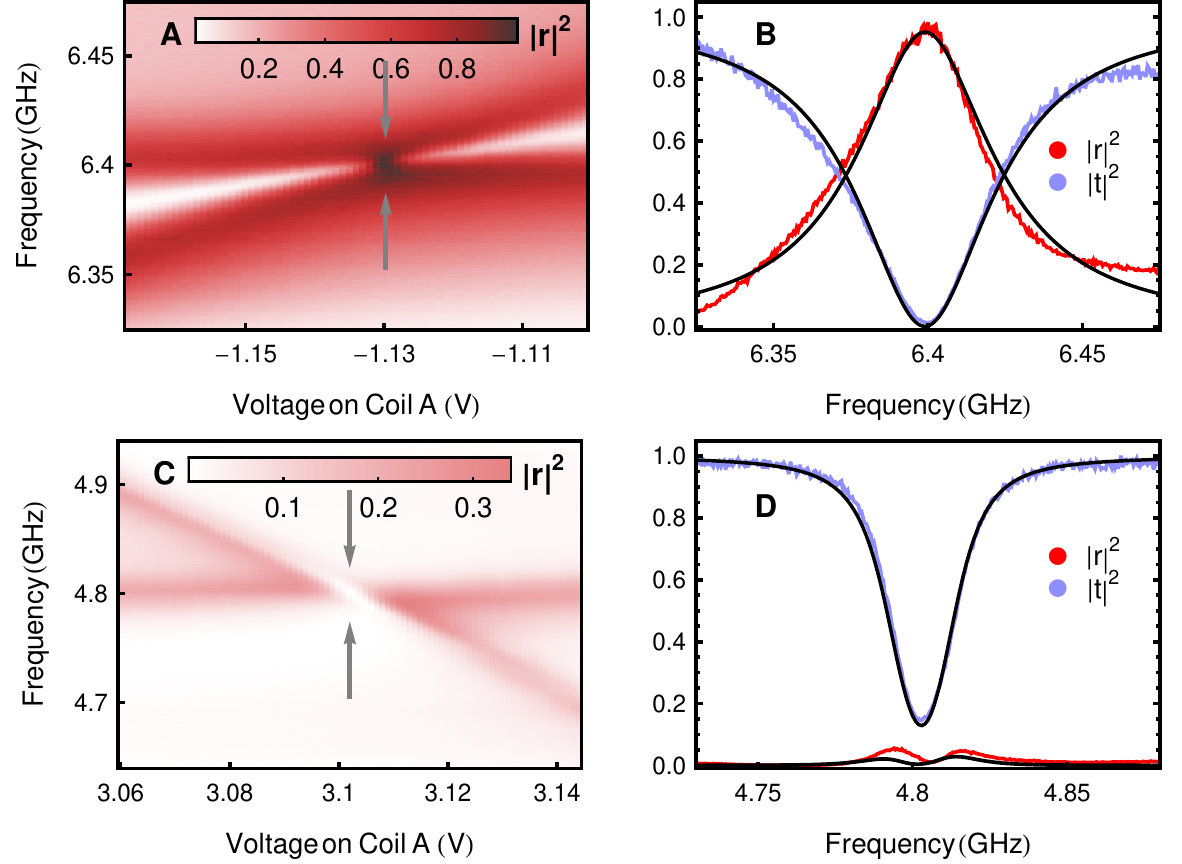}
  \end{center}
  \caption{\textbf{Elastically scattered radiation.} Reflectance spectra $|r|^2$ of the two-qubit system (A,B) at $d\sim\lambda$ recorded with a drive rate of 7.5~MHz and (C,D) at $d\sim3\lambda/4$ with rate~8.7~MHz (red data sets). In (B,D) the transmittance $|t|^2$ is also shown (blue data sets). In (A) and (C), the frequency of one qubit is tuned by applying the indicated voltages to millimeter-size coils integrated in the sample mount while the other qubit is kept at a fixed frequency. Close to resonance, interference effects cause the qubit peaks to become asymmetric. (B,D) show spectra at bias points indicated by arrows in (A,C), respectively. Colored lines are data. Black lines are theory (see text for details).}
  \label{fig:elastic}
\end{figure*}

These observations can be understood by considering photon-mediated interactions between the two qubits \cite{Lalumiere2013}. When $d\sim\lambda$, the exchange interaction $J=0$ and the correlated decay rate takes its maximal value $\gamma = \gamma_1$. This type of decay leads to subradiant, $\ket{D} = (\ket{ge} - \ket{eg} )/\sqrt{2}$, and superradiant, $\ket{B} = (\ket{ge}+\ket{eg})/\sqrt{2}$, states~\cite{DeVoe1996}. When the qubits are separated by $\lambda$, they are driven with the same phase by a single drive tone. Therefore, the symmetric state $\ket{B}$, which has the same symmetry as the drive at the qubits, can be excited from the ground state $\ket{gg}$. Thus, this state is bright. In contrast, the matrix element for the antisymmetric state $\ket{D}$, which has opposite symmetry with respect to both the drive field and the resonant vacuum fluctuations that cause relaxation, is zero. Because the latter is dark, the two-qubit system essentially behaves as a single two-level system $\{\ket{gg},\ket{B}\}$ at low drive powers with the superradiant decay rate $\Gamma_B = 2\gamma_1$~\cite{Lalumiere2013}. The two-qubit results (Fig.~2B) are indeed qualitatively similar to those of a single qubit (Fig.~1B), but with a linewidth $\Gamma_B/2\pi \sim 52 \pm 1$ MHz which is twice as large as the one extracted for a single qubit. This is a clear signature of super-radiance.

The situation is more subtle when the qubits are separated by $d\sim3\lambda/4$ or any odd multiple of $\lambda/4$. Then one of the qubits is at a node of the driving field, whereas the other is at an antinode (Fig.~1A). In this case, the above argument leading to the creation of bright and dark states does not apply and correlated decay is absent with $\gamma = 0$. Instead, the coherent exchange interaction between qubits mediated by virtual photons takes its maximal value $J = \gamma_1/2$~\footnote{Whereas correlated decay mediated by the exchange of real photons is absent when one of the qubits is at an antinode of the field, the exchange of virtual photons is still possible as it is caused by virtual photons at all frequencies except the qubit transition frequency.}.

The expected signature of this coherent exchange interaction is an anti-crossing of the qubit energy levels similar to the one observed for two qubits coupled to a resonator in a circuit QED system~\cite{Majer2007,Filipp2011a}. As the expected splitting $2J=\gamma_1$ can be only as large as the peak width $\gamma_1$, this signature is not apparent in the elastic scattering data. The doublet observed in reflection in Fig.~2D is a consequence of dressing of the two-qubit system by the input field rather than a signature of exchange interaction (see the supplementary materials). The quantitative analysis presented in \cite{Lalumiere2013} (solid black lines) agrees with the data (colored lines) over the full range of frequencies shown in Fig.~2.

To further characterize photon-mediated atom-atom interactions, the full spectrum of the elastically and inelastically scattered radiation, including the Rayleigh scattered and resonance fluorescence contributions, has been recorded in both reflection and transmission (see the supplementary materials). In a reference measurement of the resonance fluorescence spectrum of a single qubit we observe the standard Mollow triplet~\cite{Mollow1969} including a $\delta$-like peak due to elastically (Rayleigh) scattered radiation (Fig.~1C). The Mollow sidebands appear at a detuning from the center peak corresponding to the Rabi frequency induced by the drive.

When both qubits are tuned to $6.4\,\rm{GHz}$ corresponding to $d\sim\lambda$, we observe a Mollow triplet--like spectrum with a narrow resonance superimposed at its center frequency, which displays a power dependence (Fig.~3). The narrow resonance becomes more discernible as the drive power is decreased (Fig.~3B). Here and below, the Rayleigh-scattered contribution was removed from the data for clarity. These two distinct features, the Mollow triplet and narrow resonance, are due to the formation of super- and sub-radiant states, respectively.

\begin{figure}[!t]
\includegraphics[width=6.0cm]{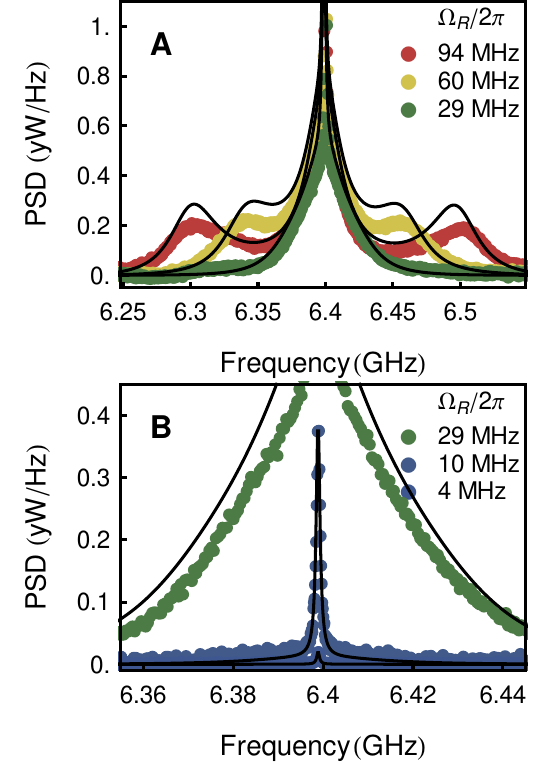}
\caption{\textbf{Super- and subradiance.} (A,B) Power spectral density measured in reflection for two qubits in resonance at $d\sim\lambda$ at the indicated Rabi drive rates $\Omega_{\rm{R}}$.
Solid lines are numerical calculations~\cite{Lalumiere2013} using the parameters specified in the supplementary materials.}
  \label{fig:supersub}
\end{figure}

The effective two-level system $\{\ket{gg},\ket{B}\}$ is strongly dressed by the drive field, resulting in a Mollow triplet. The width $\sim52~\rm{MHz}$ of the main peak, obtained by fitting to numerical calculations (black lines in Fig.~3), is consistent with the value of $\Gamma_B$ extracted above. Ideally, the dark state $\ket{D}$ is neither excited by the drive nor does it decay into $\ket{gg}$ due to selection rules. In practice, however, it is weakly populated due to qubit dephasing, non-radiative decay from the state $\ket{ee}$, and unequal single-qubit relaxation rates \cite{Lalumiere2013}. As a result, the dark state $\ket{D}$ appears as a narrow resonance superimposed on the bright-state Mollow triplet. Its linewidth, when compared to numerical results, is approximately $\Gamma_D/2\pi\sim0.4\pm0.2~\rm{MHz}$. We obtain adequate quantitative agreement between the measured spectra and theory (blakc lines in Fig.~3) and find the ratio between super- and sub-radiant lifetimes to be as high as $\Gamma_B/\Gamma_D \ge 100$.

When both qubits are tuned to $4.8\,\rm{GHz}$, corresponding to $d=3\lambda/4$, the resonance fluorescence spectra at high powers ($\Omega_{\rm{R}}/2\pi\ge 15\,\rm{MHz}$, where $\Omega_{\rm{R}}$ is the Rabi frequency) also display the expected Mollow triplet features (Fig.~4A). At drive powers much lower than the relaxation rate $\Omega_{\rm{R}}/2\pi \le 5 \,\rm{MHz}$ the fluorescence spectrum displays a double peak structure split by $\sim 15\,\rm{MHz}$ (see Fig.~4B). This observation is a clear signature of the effective exchange interaction $J$ between the two qubits mediated by virtual photons. The observed splitting is slightly larger than the expected value $2J/2\pi=\gamma_1/2\pi\approx13\,\mathrm{MHz}$, which is consistent with theory predicting the splitting to be larger than 2J in transmission and smaller in reflection \cite{Lalumiere2013}.

\begin{figure}[!t]
\includegraphics[width=6.0cm]{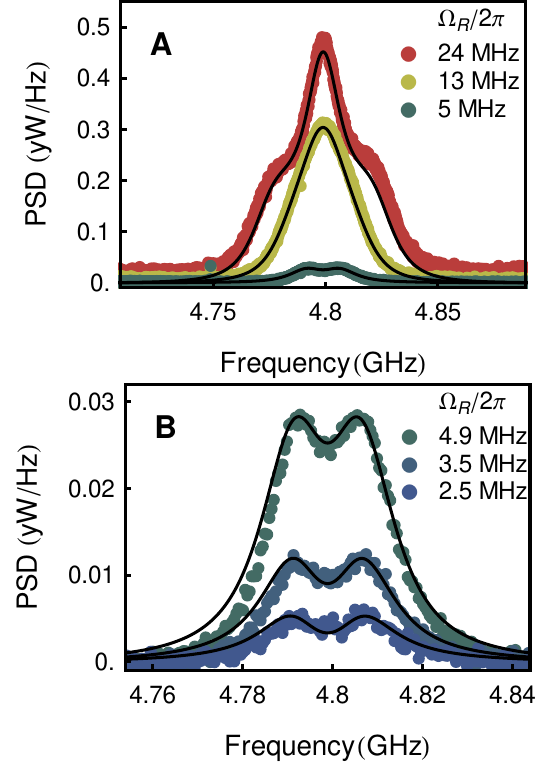}
\caption{\textbf{Exchange interaction.} (A,B) Power spectral density measured in transmission for two qubits in resonance at $d\sim3\lambda/4$ at the indicated Rabi drive rates $\Omega_{\rm{R}}$. Solid lines as in Fig.~3.}
  \label{fig:exchange}
\end{figure}
Our results present compelling evidence of strong interaction effects between two superconducting qubits separated by as much as a full wavelength $\lambda \sim 18.6\,\rm{mm}$ in an open 1D environment. The observation of these effects presents opportunities to explore decoherence-free subspaces and entanglement over long distances in open environments and could lead to exciting applications in quantum communication technology.

\begin{acknowledgments}
We acknowledge financial support by CIFAR, NSERC, AITF and ETH Zurich, and we thank Calcul Québec and Compute Canada for computational resources. Furthermore, we thank for L.~Steffen for sample fabrication, C.~Lang and Y.~Salath\'{e} for realizing the firmware used for measuring power spectral densities using an FPGA, M.~Boissonault for help with numerical simulations and V.~Sandoghdar for initial discussions motivating this work.
\end{acknowledgments}

\bibliographystyle{science}
\bibliography{refs}

\begin{thebibliography}{10}

\bibitem{DeVoe1996}
R.~G. DeVoe, R.~G. Brewer, {\it Physical Review Letters\/} {\bf 76}, 2049
  (1996).

\bibitem{Eschner2001a}
J.~Eschner, C.~Raab, F.~Schmidt-Kaler, R.~Blatt, {\it Nature\/} {\bf 413}, 495
  (2001).

\bibitem{Kimble2008}
H.~J. Kimble, {\it Nature\/} {\bf 453}, 1023 (2008).

\bibitem{Chang2007}
D.~E. Chang, A.~S. Sorensen, E.~A. Demler, M.~D. Lukin, {\it Nature Physics\/}
  {\bf 3}, 807 (2007).

\bibitem{Hwang2009}
J.~Hwang, {\it et~al.\/}, {\it Nature\/} {\bf 460}, 76 (2009).

\bibitem{Leung2012a}
P.~M. Leung, B.~C. Sanders, {\it Physical Review Letters\/} {\bf 109}, 253603
  (2012).

\bibitem{Zheng2012h}
H.~Zheng, D.~J. Gauthier, H.~U. Baranger, {\it Physical Review A\/} {\bf 85},
  043832 (2012).

\bibitem{Chang2006}
D.~E. Chang, A.~S. Sorensen, P.~R. Hemmer, M.~D. Lukin, {\it Physical Review
  Letters\/} {\bf 97} (2006).

\bibitem{Chang2007a}
D.~E. Chang, A.~S. S\o{}rensen, P.~R. Hemmer, M.~D. Lukin, {\it Physical Review
  B\/} {\bf 76}, 035420 (2007).

\bibitem{Akimov2007}
A.~V. Akimov, {\it et~al.\/}, {\it Nature\/} {\bf 450}, 402 (2007).

\bibitem{Martin-Cano2010}
D.~Mart\'{i}n-Cano, L.~Mart\'{i}n-Moreno, F.~J. Garc\'{i}a-Vidal, E.~Moreno,
  {\it Nano Letters\/} {\bf 10}, 3129 (2010).

\bibitem{Gonzalez-Tudela2011}
A.~Gonzalez-Tudela, {\it et~al.\/}, {\it Physical Review Letters\/} {\bf 106},
  020501 (2011).

\bibitem{Nayak2007}
K.~P. Nayak, {\it et~al.\/}, {\it Opt. Express\/} {\bf 15}, 5431 (2007).

\bibitem{Vetsch2010}
E.~Vetsch, {\it et~al.\/}, {\it Physical Review Letters\/} {\bf 104}, 203603
  (2010).

\bibitem{Chang2012a}
D.~E. Chang, L.~Jiang, A.~V. Gorshkov, H.~J. Kimble, {\it New Journal of
  Physics\/} {\bf 14}, 063003 (2012).

\bibitem{Goban2012}
A.~Goban, {\it et~al.\/}, {\it Physical Review Letters\/} {\bf 109}, 033603
  (2012).

\bibitem{Shen2005}
J.-T. Shen, S.~Fan, {\it Physical Review Letters\/} {\bf 95}, 213001 (2005).

\bibitem{Astafiev2010}
O.~Astafiev, {\it et~al.\/}, {\it Science\/} {\bf 327}, 840 (2010).

\bibitem{Abdumalikov2010}
A.~A. Abdumalikov, {\it et~al.\/}, {\it Physical Review Letters\/} {\bf 104},
  193601 (2010).

\bibitem{Abdumalikov2011}
A.~A. Abdumalikov, O.~V. Astafiev, Y.~A. Pashkin, Y.~Nakamura, J.~S. Tsai, {\it
  Physical Review Letters\/} {\bf 107}, 043604 (2011).

\bibitem{Hoi2013b}
I.-C. Hoi, {\it et~al.\/}, {\it Physical Review Letters\/} {\bf 111}, 053601
  (2013).

\bibitem{Hoi2011}
I.-C. Hoi, {\it et~al.\/}, {\it Physical Review Letters\/} {\bf 107}, 073601
  (2011).

\bibitem{Lalumiere2013}
K.~Lalumi\`{e}re, {\it et~al.\/}, {\it Physical Review A\/}  (2013).

\bibitem{Koch2007}
J.~Koch, {\it et~al.\/}, {\it Physical Review A\/} {\bf 76}, 042319 (2007).

\bibitem{Hoi2012b}
I.-C. Hoi, {\it et~al.\/}, {\it Physical Review Letters\/} {\bf 108}, 263601
  (2012).

\bibitem{Note1}
Whereas correlated decay mediated by the exchange of real photons is absent
  when one of the qubits is at an antinode of the field, the exchange of
  virtual photons is still possible as it is caused by virtual photons at all
  frequencies except the qubit transition frequency.

\bibitem{Majer2007}
J.~Majer, {\it et~al.\/}, {\it Nature\/} {\bf 449}, 443 (2007).

\bibitem{Filipp2011a}
S.~Filipp, {\it et~al.\/}, {\it Physical Review A\/} {\bf 83}, 063827 (2011).

\bibitem{Mollow1969}
B.~R. Mollow, {\it Physical Review\/} {\bf 188}, 1969 (1969).

\bibitem{Lang2011}
C.~Lang, {\it et~al.\/}, {\it Physical Review Letters\/} {\bf 106}, 243601
  (2011).

\end{thebibliography}

\appendix
\newpage

\section{Supplementary material}

\subsection{Measuring elastically scattered radiation}
\label{app:elastic}
The microwave tone transmitted through or reflected from the sample is amplified and then down-converted to $25\,\rm{MHz}$ using an analog microwave frequency mixer. It is then digitized at a rate of $100$ MHz. Both amplitude and phase of the fields are extracted by digital down-conversion. Typically, $1.0\,10^3$ to $3.3\,10^{4}$ traces, each of duration $8192\,\rm{ns}$ are averaged in order to reach the signal-to-noise ratio displayed in Figs.~1 and 2.

\subsection{Background subtraction}
\label{app:background}
To compensate for the undesired frequency-dependent transmission and reflection properties of the drive and detection lines, we first measure the frequency dependence of the background transmission and reflection coefficients ($t_{bg}$ and $r_{bg}$) with the qubits tuned out of the frequency range of interest. The coefficients $t_p$ and $r_p$ are then measured with the qubits tuned to the desired frequencies. The transmittance $|t|^2$ and reflectance $|r|^2$ are then calculated according to $|t|^2=|t_{p}|^2/(|t_{bg}|^2+|r_{bg}|^2)$ and $|r|^2=|r_{p}|^2/(|t_{bg}|^2+|r_{bg}|^2)$.

The background subtraction method is of limited accuracy due to standing waves which form between the impedance discontinuities induced by the qubits at their given frequencies and impedance mismatches in the feed and detection lines. These effects lead to a small modulation of the transmittance and reflectance data as observed for example in Fig.~1B and Fig.~2B,D.

\subsection{Strong coupling}
\label{app:strongcoupling}
The transmission coefficient at resonance and low drive power is given by $t_{\rm{min}}=1 - (\gamma_1/(\gamma_1 + \gamma_{\mathrm{nr}} + 2 \gamma_{\varphi}))$ where $\gamma_1$ is the rate at which the atom emits into the transmission line, $\gamma_{\mathrm{nr}}$ denotes radiation into other channels, and $\gamma_{\varphi}$ is the pure dephasing rate. In the absence of non-radiative decay and pure dephasing, the transmission will go to zero. In Fig.~1B, the transmittance $|t|^2$ was observed to be lower than 0.025 at resonance, indicating that $t_{min}<0.16$. The combination of non-radiative scattering and dephasing leading to less than $16\%$ loss indicates that the system is in the strong coupling regime of waveguide QED.

\subsection{Resonance fluorescence measurements}
\label{app:resfluor}
To record the full spectrum, the down-converted field amplitudes were measured as a function of time for $8192$ ns at a sampling rate of 1~GHz. Each time trace was Fourier transformed and multiplied with its complex conjugate using fast digital electronics to obtain the power spectral density $S(\omega)$ of the detected radiation~\cite{Lang2011}. The resulting data was averaged between $3.4\,10^{7}$ and $1.0\,10^9$ times to achieve the signal to noise ratio displayed in the figures.

\subsection{Qubit saturation}
\label{app:saturation}
In Fig.~2B, the transmittance and reflectance approximately add to unity ($|r|^2+|t|^2\approx1$) indicating that little or no radiation is inelastically scattered at the given drive strength $\Omega_R \approx 7.5\,\rm{MHz}$ and  relaxation rate $\gamma_1 \approx 26\,\rm{MHz}$. In contrast in Fig.~2D in agreement with theory (solid lines) $|r|^2+|t|^2$ deviates strongly from unity for $\Omega_R \approx 8.7\,\rm{MHz}$ and qubit relaxation rate $\gamma_1 \approx 13\,\rm{MHz}$. This can be understood as due to the increased inelastic scattering~\cite{Astafiev2010} into frequencies outside of our detection bandwidth and a higher pure dephasing rate at this qubit transition frequency.

\subsection{Theoretical model and fits}\label{app:fits}
The model used for the fits is based on an effective 2-qubit master equation obtained when tracing out all modes of the continuum~\cite{Lalumiere2013}. In this model, qubit-qubit interaction emerges as a second order effect from the interactions between qubits and all photon modes of the continuum. The fits presented in this work are all based on a single set of parameters stated below for both qubits. The distance between the qubits was inferred to be equal to $\lambda$ at $\sim$ 6.4~GHz from the oscillatory behaviour of $\gamma$ and $J$. The maximum frequencies and anharmonicities for each qubit were measured using spectroscopic techniques. From these the Josephson and charging energies of the qubit were determined~\cite{Koch2007}. The dimensionless coupling strength $g_j$ between qubit $j$ and the line is defined as \cite{Lalumiere2013}
\begin{equation}
  g_j = \sqrt{ \frac{e^2 c_t}{2 \hbar \pi v c_{g,j}^2}} \left( \frac{E_{J}}{8E_{C}}\right)^{{1}/{4}}
  \label{gk}
\end{equation}
with $c_t$ the capacitance per length of the transmission line, $v$ the speed of light in the line, and $c_{g,j}$ the capacitance between qubit $j$ and the transmission line. From spectroscopic measurements  we find that $g_1=0.0146$ and $g_2=0.0180$. The nonradiative decay rates $\gamma_{\mathrm{nr}}/2\pi=$ 1.2~MHz and pure dephasing rates $\gamma_{\varphi}/2\pi=$ 1.8~MHz are assumed to be identical for both qubits at 4.8~GHz. At 6.4~GHz, $\gamma_{\varphi}/2\pi=$ 0.2~MHz, consistent with our expectation that pure dephasing is more prominent at lower qubit frequencies.
At this frequency $\gamma_{\mathrm{nr}}/2\pi$ varies from 0.18~MHz to 0.58~MHz with drive power. 
Additionally, a small offset was used as a free parameter when fitting to the experimentally obtained power spectral densities. 

\end{document}